\pdfoutput=1
\documentclass[prd,nofootinbib,twocolumn,superscriptaddress]{revtex4-1}

\usepackage{amsmath}
\usepackage{euscript,amssymb,epsfig}
\usepackage{amsfonts,latexsym}
\usepackage{color}

\newcommand{\be}[1]{\begin{equation}\label{#1}}
\newcommand{\ee}{\end{equation}}
\newcommand{\ba}[1]{\begin{eqnarray}\label{#1}}
\newcommand{\ea}{\end{eqnarray}}
\newcommand{\rf}[1]{(\ref{#1})}
\newcommand{\nn}{\nonumber}

\newcommand{\const}{\mbox{\rm const}\,}

\begin{document}

\title{Scalar perturbations in cosmological $f(R)$ models: the cosmic screening approach}

\author{\"{O}zg\"{u}r Akarsu}
\email{akarsuo@itu.edu.tr}
\affiliation{Department of Physics, Istanbul Technical University, Maslak 34469 Istanbul, Turkey}

\author{Ruslan Brilenkov}
\email{ruslan.brilenkov@gmail.com}
\affiliation{Institute for Astro- and Particle Physics, University of Innsbruck, Technikerstrasse 25/8, A-6020 Innsbruck, Austria}
\affiliation{Dipartimento di Fisica, Universit\`{a} di Roma `Tor Vergata', via della Ricerca Scientifica 1, 00133 Roma, Italy}
\affiliation{Institut f{\"u}r Astrophysik, Georg-August-Universit{\"a}t G{\"o}ttingen, Friedrich-Hund-Platz 1, 37077 G{\"o}ttingen, Germany}

\author{Maxim Eingorn}
\email{maxim.eingorn@gmail.com}
\affiliation{Department of Mathematics and Physics, North Carolina Central University, Fayetteville st. 1801, Durham, North Carolina 27707, U.S.A.}

\author{Valerii Shulga}
\email{shulga@rian.kharkov.ua}
\affiliation{International Center of Future Science of the Jilin University, 2699 Qianjin St., 130012, Changchun City, China}
\affiliation{Institute of Radio Astronomy of National Academy of Sciences of Ukraine, 4 Mystetstv str., 61002 Kharkiv, Ukraine}

\author{Alexander Zhuk}
\email{ai.zhuk2@gmail.com}
\affiliation{Department of Physics, Istanbul Technical University, Maslak 34469 Istanbul, Turkey}
\affiliation{Astronomical Observatory, Odessa National University, Dvoryanskaya st. 2, 65082 Odessa, Ukraine}

\begin{abstract}
We investigate cosmological perturbations for nonlinear $f(R)$ models within the cosmic screening approach. Matter is considered both in the form of a set of
discrete point-like massive bodies and in the form of a continuous pressureless perfect fluid. We perform full relativistic analysis of the first-order theory of
scalar perturbations for arbitrary nonlinear $f(R)$ models and demonstrate that scalar potentials $\Phi(t,\mathbf{r})$ and $\Psi(t,\mathbf{r})$ are determined by
a system of only two master equations. Our equations are applicable at all spatial scales as long as the approximation $\delta R/\bar R \ll 1$ (which is usually
assumed in studies devoted to cosmological perturbations in $f(R)$ models) works.

\end{abstract}
\date{\today}

%\keywords{}
\maketitle

\section{Introduction}

\setcounter{equation}{0}

Since the discovery of the late time accelerated expansion of our Universe at the end of the last century, the mystery of dark energy remains one of the most
intriguing subjects in modern cosmology. One of the possible solutions, alternative to the introduction of dark energy, is to modify gravity, namely, the general
theory of relativity. This idea was known from the eighties of the last century, when $R$ of Einstein-Hilbert action replaced by $R+\alpha R^2$ of Starobinsky
action was considered to describe the accelerated expansion of the early Universe (so called inflation) \cite{Starobinsky:1980te} \footnote{One can also see
\cite{Saidov:2010wx} for a possibility of inflation from quadratic and quartic nonlinearities in multidimensional cosmological models.}. It was then realized that
nonlinear $f(R)$ models can also describe the late acceleration. This resulted in a huge number of articles devoted to the study of nonlinear gravitational models
(see, e.g., reviews \cite{Sotiriou:2008rp,DeFelice:2010aj,Nojiri:2010wj,Motohashi:2010zz,Motohashi:2012wh,Koyama:2015vza,Nojiri:2017ncd,Burrage:2017qrf}). Because
the nature of dark energy is not clear yet, these theories are still of unflagging interest (see, e.g., the most recent papers
\cite{Capozziello:2018oiw,Peel:2018aly,Lopes:2018uhq,Roncarelli:2018kud}).

Obviously, the observable large scale structure of the Universe is the crucial test of any gravitational theory. Comparing the predictions of such a theory with
the observational picture, we can conclude how viable it is. This can be done within the framework of perturbation theory. Usually, the Friedmann Universe is
considered as a background cosmological model. For such background in the case of the first-order theory of scalar perturbations, the metric perturbations are
characterized by two functions $\Phi$ and $\Psi$ \cite{Mukhanov:1990me,Gorbunov:2011zzc}, where the former one is the gravitational potential created by
inhomogeneities.
%We need to know these functions to perform numerical simulation of the large scale structure.
For $f(R)$ models, the system of equations for $\Phi$ and $\Psi$ was obtained in \cite{Hwang:2001qk,Hwang:2005hb} (see also \cite{DeFelice:2010aj}). The energy
density fluctuation $\delta\varepsilon =-\delta T^0_0$ is the source of the potentials $\Phi$ and $\Psi$ in this system\footnote{In the case of linear gravity
(i.e. $f(R)=R$) with matter in the form of a perfect fluid, the potentials are equal to each other: $\Phi=\Psi$.}. However, since the energy-momentum tensor
$T^{\alpha\beta}$ depends on metric, the energy density fluctuation is a function of potentials $\Phi$ and $\Psi$. For example, in the linear gravity model the
term proportional to $\bar\varepsilon\Phi$ contributes to $\delta\varepsilon$ (see also Eq.~\rf{2.17} below), where $\bar\varepsilon$ is the averaged energy
density. In the literature which we are aware of and which is devoted to the study of perturbations in nonlinear models (see, for example, reviews \cite{Sotiriou:2008rp,DeFelice:2010aj,Nojiri:2010wj,Motohashi:2010zz,Motohashi:2012wh,Koyama:2015vza,Nojiri:2017ncd,Burrage:2017qrf} and references therein, in particular, the paper \cite{delaCruzDombriz:2008cp}), the explicit dependence of the energy-momentum tensor on the perturbations of the metric is not taken into account. This is one of the main differences between our approach and these articles. This drastically changes the form of the gravitational interaction.  As it was shown in our papers
\cite{Eingorn:2015hza,Eingorn:2016kdt,Eingorn:2015yra,Eingorn:2017}, the gravitational potential satisfies the Helmholtz-type equation but not the Poisson
equation. Therefore, it has the form of the Yukawa potential. It is important to note that the cosmological background in the form of the Friedmann Universe with
the nonzero averaged energy density $\bar\varepsilon\neq 0$ plays the crucial role in this effect. In the case of the Minkowski background $\bar\varepsilon=0$,
and we go back to the Poisson equation which has a solution in the form of Newtonian gravitational potential in full agreement with the textbook \cite{Landau}.
Since the gravitational potential of an individual inhomogeneity is exponentially suppressed at large cosmological scales, we call this effect a cosmic screening
\cite{Eingorn:2016kdt,Eingorn:2017adg} \footnote{In our previous  mechanical approach \cite{Eingorn:2012jm,Eingorn:2013faa,Eingorn:2012dg,Zhuk:2016imt}, we also
took into account the dependence of the perturbed energy density on the gravitational potential. However, since %for some erroneous reasons
we dropped the additional term $\bar\varepsilon \Phi$ in the perturbed Einstein equation for $\Phi$, the exponential suppression of gravitational interaction was
not observed in this approach. The theory of scalar perturbations for nonlinear $f(R)$ models within the mechanical approach was studied in the paper
\cite{Eingorn:2014laa}.
%In the cosmic screening approach, in contrast to the
%mechanical approach, the peculiar velocities and the vector perturbations are taken into account.
For the linear $f(R)=R$ model, the second-order cosmological perturbations within the cosmic screening approach were studied in \cite{Brilenkov:2017gro}.}. It is
worth noting that nonlinearity of gravity in $f(R)$ models also affects the form of the gravitational potential \cite{DeFelice:2010aj}. For example, at the
astrophysical scales the Newtonian gravitational potential acquires an additional term in the form of the Yukawa potential \cite{Eingorn:2011kx,Eingorn:2011aa}.

In the present paper we study the first-order scalar cosmological perturbations for nonlinear $f(R)$ models within the cosmic screening approach. We demonstrate
that in this case a rather complicated initial system of six differential equations is reduced to two equations for the potentials $\Phi$ and $\Psi$. These
equations establish the basis for numerical simulation of the large-scale structure of the Universe in nonlinear gravity models.

The paper is structured as follows. In Sec. 2, we describe the model and present the basic equations at the background and perturbation levels. In Sec. 3, we
obtain two master equations for the potentials $\Phi$ and $\Psi$. The main results are briefly summarized in the concluding Sec. 4.

\section{The model and basic equations}

\setcounter{equation}{0}

In this section we present the basic equations that we will use hereinafter. For background and perturbed equations we follow mainly the review
\cite{DeFelice:2010aj} using the sign convention accepted in this paper. In $f(R)$ gravity, the action reads
%%%%%%
\be{2.1} S=\frac{1}{2\kappa^2}\int d^4x\sqrt{-g}f(R) + S_m\, . \ee
%%%%%%
Here $f(R)$ is an arbitrary function of the scalar curvature $R$, $S_m$ is the action for matter, $\kappa^2\equiv 8\pi G_N$, where $G_N$ stands for Newtonian
gravitational constant, and we use units such that the speed of light is equal to 1:  $c\equiv 1$. The equation of motion corresponding to this action is
%%%%%%
\ba{2.2} &{}&F(R)R_{\mu\nu} - \frac{1}{2}f(R)g_{\mu\nu} - \nabla_{\mu}\nabla_{\nu}F(R) + g_{\mu\nu}\square F(R)\nn\\
&=&{\kappa}^2 T_{\mu\nu}\, \quad \mu,\nu=0,1,2,3\, . \ea
%%%%%%
The trace of this equation gives
%%%%%%
\be{2.3}
3\square F(R)+F(R)R-2f(R)=\kappa^2 T\, ,
\ee
%%%%%%
where $T=g^{\mu\nu}T_{\mu\nu}$ and $F(R)\equiv f'(R)$.  Hereinafter, the prime denotes the derivative with respect to the scalar curvature $R$, i.e. $f'(R)\equiv
d f/d R$. Since $f(R)$ has dimension of $R$, the function $F(R)$ is dimensionless. Besides, $\square F =
(1/\sqrt{-g})\partial_{\mu}(\sqrt{-g}g^{\mu\nu}\partial_{\nu}F)$.
%Greek indices $\mu,\nu=0,1,2,3$, the Latin indices $i,j=1,2,3$ and

In the case of the spatially flat background spacetime with Friedmann-Robertson-Walker (FRW) metric
%%%%%
\be{2.4}
ds^2=g_{\mu\nu}dx^{\mu}dx^{\nu}=-dt^2+a^2(t) \left(dx^2+dy^2+dz^2\right)
\ee
%%%%%
and matter in the form of a perfect fluid with the energy-momentum tensor components $\bar T^{\mu}_{\nu}=\mathrm{diag}(-\bar\varepsilon,\bar P,\bar P,\bar P)$,
Eq.~\rf{2.2} results in the following system of background equations:
%%%%%
\be{2.5} 3FH^2 = (FR - f)/2 - 3H\dot{F} + \kappa^2 \bar \varepsilon \, \ee
%%%%%%
and
%%%%%%
\be{2.6} -2F\dot{H}= \ddot {F} - H\dot{F} + \kappa^2 (\bar\varepsilon + \bar P)\, . \ee
%%%%%
Here the bar denotes homogeneous background quantities, the Hubble parameter $H\equiv\dot a/a$ (the dot everywhere denotes the derivative with respect to the
synchronous time $t$) and the scalar curvature
%%%%%
\be{2.7}
R=6\left(2H^2 + \dot{H}\right)\, .
\ee
%%%%%%
The perfect fluid  satisfies the continuity equation
%%%%%
\be{2.8} \dot{\bar\varepsilon} + 3H(\bar\varepsilon + \bar P)=0\, , \ee
%%%%%%
which for nonrelativistic matter with $P=0$ has the solution
%%%%%%
\be{2.9} \bar \varepsilon = \bar\rho/a^3\, , \ee
%%%%%%
where $\bar\rho=\const$ is the averaged mass density in the comoving coordinates.
%Since the speed of light $c=1$, then in this units $\bar\rho =\bar\varepsilon$, where $\bar\varepsilon$ is the averaged energy density.
In what follows we will consider matter either in the form of a set of discrete point-like massive bodies or in the form of a continuous pressureless perfect
fluid. For both of these cases the pressure and its fluctuation are equal to zero: $P=\delta P=0$.

\

Let us turn now to the perturbed equations. Inhomogeneities or fluctuations of matter result in perturbations of the FRW metric \rf{2.4}. In our paper we restrict
ourselves to the scalar perturbations. In the conformal Newtonian (longitudinal) gauge, the perturbed metric reads \cite{Mukhanov:1990me,Gorbunov:2011zzc}
%%%%%%%
\be{2.10}
ds^2= -(1+2\Phi)dt^2+a^2(1-2\Psi)\left(dx^2+dy^2+dz^2\right)\, ,
\ee
%%%%%%%
where the introduced scalar perturbations $\Phi,\Psi \ll 1$ depend on all spacetime coordinates. These perturbations satisfy the following system of linearized
equations \cite{DeFelice:2010aj}:
%%%%%%
\ba{1}
&-&\frac{\triangle\Psi}{a^2}+3H\left(H\Phi+\dot\Psi\right)\nn\\
&=&-\frac{1}{2F}\left[\left(3H^2+3\dot{H}+\frac{\triangle}{a^2}\right)\delta F \right. \nn\\
&-&\left.3H\dot{\delta F} + 3H\dot{F}\Phi + 3\dot{F}\left(H\Phi+\dot{\Psi}\right) + \kappa^2\delta\varepsilon\right]\, , \ea
%%%%%%
%%%%%%
\be{2} H\Phi+\dot{\Psi}=\frac{1}{2F}\left(\dot{\delta F}-H\delta F-\dot{F}\Phi-\frac{\kappa^2}{a^2}\Xi\right)\, , \ee
%%%%%%
%%%%%
\be{3} -F\left(\Phi-\Psi\right)=\delta F\, , \ee
%%%%%
%%%%%
\ba{4}
&{}&3\left(\dot{H}\Phi+H\dot{\Phi}+\ddot{\Psi}\right)+6H\left(H\Phi+\dot{\Psi}\right)+3\dot{H}\Phi+\frac{\triangle\Phi}{a^2}\nn\\
&=& \frac{1}{2F}\left[3\ddot{\delta F}+3H\dot{\delta F}-6H^2\delta F-\cfrac{\triangle\delta F}{a^2}-3\dot{F}\dot{\Phi}\right.\nn\\
&-&\left.3\dot{F}\left(H\Phi+\dot\Psi\right)-\left(3H\dot{F}+6\ddot{F}\right)\Phi+\kappa^2\delta\varepsilon\right]\, , \ea
%%%%%
%%%%%
\ba{5}
&{}&\ddot{\delta F}+3H\dot{\delta F}-\cfrac{\triangle\delta F}{a^2}-\frac{1}{3}R\delta F = \frac{1}{3}\kappa^2\delta\varepsilon+2\ddot{F}\Phi\nn\\
&+&\dot{F}\left(3H\Phi+3\dot{\Psi}+\dot{\Phi}\right)+3H\dot{F}\Phi-\frac{1}{3}F\delta R\, , \ea
%%%%%
%%%%%
\ba{6} &{}&\delta F=F'\delta R, \quad \delta R=-2\left[3\left(\dot{H}\Phi+H\dot\Phi+\ddot{\Psi}\right) \right.\nn\\
&+& \left.12H\left(H\Phi+\dot{\Psi}\right) +\frac{\triangle
\Phi}{a^2}+3\dot{H}\Phi-2\frac{\triangle\Psi}{a^2} \right]\, . \ea
%%%%%
In these equations the function $F$, its derivative $F'$ and the scalar curvature $R$ are unperturbed background quantities, i.e. here $F(R)\equiv F(\bar R)$.
%To simplify the form of equations we drop the sign bar over these functions that should not cause any confusion.
This system of equations is usually used in papers devoted to cosmological perturbations in $f(R)$ models. It is important to note here that the relation $\delta
F=F(R)-F(\bar R)=F'(\bar R)\delta R$ is valid only in approximation $\delta R/\bar R \ll 1$.
%For such approximation the following relation holds: $\overline{F(R)}=F(\bar R)$.
However, this approximation is violated at the late stage of the evolution of the Universe near inhomogeneities, i.e. in the vicinity of galaxies. Therefore the
system of Eqs. \rf{1}-\rf{6} and, consequently, the solutions of this system are only applicable at sufficiently large scales in the late Universe or at all
scales in the early Universe.
%when the hydrodynamic approach is still valid.

The energy density fluctuation $\delta \varepsilon$ reads
%%%%%
\be{2.17} \delta\varepsilon = \frac{\delta \rho (t,\mathbf{r})}{a^3} +\frac{3\bar \rho}{a^3}\Psi\, , \ee
%%%%%
where $\mathbf{r}$ is a comoving radius-vector. For the linear $f(R)=R$ model the potentials $\Psi$ and $\Phi$ are equal to each other: $\Psi= \Phi$. Then,
Eq.~\rf{2.17} reproduces the result of papers \cite{Eingorn:2015hza,Eingorn:2016kdt,Eingorn:2015yra}. In the case of inhomogeneities in the form of discrete
point-like  masses $m_p$, this equation can be directly obtained from the energy-momentum tensor \cite{Landau}
%%%%%%
\be{2.18} T^{ik} =\sum\limits_p\frac{m_p}{\sqrt{-g}}\frac{dx_p^i}{dt}\frac{dx_p^k}{dt}\frac{1}{d\tau_p/dt}\delta({\bf r}-{\bf r}_p)\, , \ee
%%%%%%%
where $d\tau^2=-ds^2$ and the metric is given by \rf{2.10}. The fluctuation of the mass density is
%%%%%%
\be{2.19}
\delta\rho=\rho -\bar\rho = \sum_p m_p \delta({\bf r}-{\bf r}_p) -\bar\rho\, .
\ee
%%%%%%
In the case of a continuous pressureless perfect fluid with the mass density $\rho (t,\mathbf{r})$, Eq.~\rf{2.18} is generalized as follows:
%%%%%
\be{2.20}
T^{ik}=\frac{\rho}{\sqrt{-g}} \frac{dx^i}{d\tau}\frac{dx^k}{d\tau}\frac{d\tau}{dt}=\varepsilon u^i u^k\, ,
\ee
where
%%%%%
\be{2.21} \varepsilon\equiv \frac{\rho}{\sqrt{-g}} \frac{d\tau}{dt}\, ,\quad u^i\equiv\frac{dx^i}{d\tau}\, . \ee
%%%%%%
It can be easily seen that for the energy density fluctuation we reproduce Eq.~\rf{2.17}.

The quantity $\Xi$ describes the effective peculiar velocity potential \cite{Eingorn:2015hza,Eingorn:2016kdt}. For example, in the case of discrete sources we
have
%%%%%%
\be{2.22}
a\nabla \left(\sum_p \rho_p {\mathbf{v}}_p\right)=\triangle \Xi\, ,
\ee
%%%%%%
where $\rho_p\equiv m_p\delta (\mathbf{r}-\mathbf{r}_p)$ and ${v}^{\alpha}_p\equiv dx^{\alpha}_p/dt, \, \alpha=1,2,3$. This equation can be solved exactly
\cite{Eingorn:2015hza}:
%%%%%%
\be{2.23}
\Xi = \frac{a}{4\pi}\sum_p m_p\frac{(\mathbf{r}-\mathbf{r}_p){\mathbf{v}}_p}{|\mathbf{r}-\mathbf{r}_p|^3}\, .
\ee
%%%%%%
For the continuous perfect fluid with the energy density \rf{2.21} the function $\Xi$ satisfies the equation
%%%%%
\be{2.24} a\nabla \left( \rho \mathbf{v}\right)=\triangle \Xi\, . \ee
%%%%%

\section{Master equations for the potentials $\Phi$ and $\Psi$}

\setcounter{equation}{0}

We have six equations \rf{1}--\rf{6} for two functions $\Phi(t,\mathbf{r})$ and $\Psi(t,\mathbf{r})$. The main goal now is to maximally simplify this system by
reducing the number of equations. As will be seen in what follows, we obtain a system of only two independent differential equations, which we call the master
equations for $\Phi$ and $\Psi$.

First, substituting \rf{3} into Eqs. \rf{1}, \rf{2}, \rf{4}, \rf{5} and \rf{6}, we obtain respectively:
%%%%%%
\ba{7} &-&\frac{\triangle\Psi}{a^2}+3H\left(H\Phi+\dot\Psi\right) \nn\\
&=& -\frac{1}{2F}\left[-\left(3H^2+3\dot{H}+\frac{\triangle}{a^2}\right)F\left(\Phi-\Psi\right)
\right.\nn \\
&+&3H\left(\dot{F}\left(\Phi-\Psi\right)+F\left(\dot{\Phi}-\dot{\Psi}\right)\right)+3H\dot{F}\Phi \nn\\
&+&\left. 3\dot{F}\left(H\Phi+\dot{\Psi}\right) +
\kappa^2\delta\varepsilon\right]\, , \ea
%%%%%%

\vspace{1mm}

%%%%%%
\ba{8} &{}&H\Phi+\dot{\Psi}=\frac{1}{2F}\left[-\left(\dot{F}\left(\Phi-\Psi\right)+F\left(\dot{\Phi}-\dot{\Psi}\right)\right)\right.\nn\\
&+&\left.HF\left(\Phi-\Psi\right)-\dot{F}\Phi-\frac{\kappa^2}{a^2}\Xi\right]\, , \ea
%%%%%%

\vspace{1mm}

%%%%%
\ba{9}
&{}&3\left(\dot{H}\Phi+H\dot{\Phi}+\ddot{\Psi}\right)+6H\left(H\Phi+\dot{\Psi}\right)+3\dot{H}\Phi+\frac{\triangle\Phi}{a^2}\nn\\
&=&\frac{1}{2F}\left[-3\left(\ddot{F}\left(\Phi-\Psi\right)+2\dot{F}\left(\dot{\Phi}-\dot{\Psi}\right)+F\left(\ddot{\Phi}-
\ddot{\Psi}\right)\right)\right.\nn\\
&-&3H\left(\dot{F}\left(\Phi-\Psi\right)+F\left(\dot{\Phi}-\dot{\Psi}\right)\right)+6H^2F\left(\Phi-\Psi\right)\nn\\
&+&\cfrac{F}{a^2}\left(\triangle\Phi-
\triangle\Psi\right) - 3\dot{F}\dot{\Phi}-3\dot{F}\left(H\Phi+\dot\Psi\right)\nn\\
&-&\left.\left(3H\dot{F}+6\ddot{F}\right)\Phi+\kappa^2\delta\varepsilon\right] \, , \ea
%%%%%

\vspace{1mm}

%%%%%
\ba{10} &-&\left(\ddot{F}\left(\Phi-\Psi\right)+2\dot{F}\left(\dot{\Phi}-\dot{\Psi}\right)+F\left(\ddot{\Phi}-\ddot{\Psi}\right)\right)\nn\\
&-&3H\left(\dot{F}\left(\Phi-\Psi\right)+F\left(\dot{\Phi}-\dot{\Psi}\right)\right)+\cfrac{F}{a^2}\left(\triangle\Phi-\triangle\Psi\right)\nn\\
&+&\frac{1}{3}RF\left(\Phi-\Psi\right)= \frac{1}{3}\kappa^2\delta\varepsilon
+\dot{F}\left(3H\Phi+3\dot{\Psi}+\dot{\Phi}\right) \nn\\
&+&2\ddot{F}\Phi+3H\dot{F}\Phi +\frac{2}{3}F\left[3\left(\dot{H}\Phi+H\dot\Phi+\ddot{\Psi}\right) \right.\nn\\
&+&\left.12H\left(H\Phi+\dot{\Psi}\right)+ \frac{\triangle \Phi}{a^2}+3\dot{H}\Phi-2\frac{\triangle\Psi}{a^2} \right] \, , \ea
%%%%%

\vspace{1mm}

%%%%%
\ba{11} &-&F\left(\Phi-\Psi\right)=-2F'\left[3\left(\dot{H}\Phi+H\dot\Phi+\ddot{\Psi}\right)\right.\nn\\
&+&\left.12H\left(H\Phi+\dot{\Psi}\right)+
\frac{\triangle\Phi}{a^2}+3\dot{H}\Phi-2\frac{\triangle\Psi}{a^2} \right]\, .\ea
%%%%%
%%%%%%%%

Next, from Eqs. \rf{10} and \rf{11} we find:
%%%%%
\ba{15} \ddot{\Psi}&=&\frac{F}{6F'}\left(\Phi-\Psi\right)-4H\left(H\Phi+\dot{\Psi}\right)-
\frac{1}{3}\frac{\triangle\Phi}{a^2}-2\dot{H}\Phi\nn\\
&+&\frac{2}{3}\frac{\triangle\Psi}{a^2}-H\dot\Phi\, \ea
%%%%%
and
%%%%%
\ba{18}\ddot{\Phi}&=&\dot{\Phi}\left[-3\frac{\dot{F}}{F}-4H\right]-\dot{\Psi}\left[\frac{\dot{F}}{F}+H\right]
+\Phi\left[-3\frac{\ddot{F}}{F} -2\dot{H}\right.\nn\\
&-&\left.9H\frac{\dot{F}}{F}
+\frac{1}{3}R-4H^2-\frac{F}{6F'}\right] +\cfrac{1}{3a^2}\left(2\triangle\Phi-\triangle\Psi\right)
\nn\\
&-&\Psi\left[-\frac{\ddot{F}}{F}-3H\frac{\dot{F}}{F}+\frac{1}{3}R-\frac{F}{6F'}\right]
-\frac{1}{3F}\kappa^2\delta\varepsilon\, . \ea
%%%%%
Substituting these expressions for $\ddot\Phi$ and $\ddot\Psi$ into Eq.~\rf{9}, after rather tedious but not complicated calculations we get
%%%%%
\ba{23} &{}&6HF\dot{\Phi}+\dot{\Psi}\left[6\dot{F}+6HF\right]\nn\\
&+&\Phi\left[18H\dot{F}-FR+18FH^2\right]\nn\\
&+&\Psi\left[-6H\dot{F}+FR-6H^2F\right]\nn\\
&-&2\cfrac{F}{a^2}\left(\triangle\Phi+\triangle\Psi\right)+2\kappa^2\delta\varepsilon=0\, . \ea
%%%%%
Eq.~\rf{8} can be written in the form
%%%%%%
\ba{26} F\dot{\Phi}&=&-F\dot{\Psi}-\Phi\left[2\dot{F}+FH\right]\nn\\
&-&\Psi\left[-\dot{F}+FH\right]-\frac{\kappa^2}{a^2}\Xi\, . \ea
%%%%%%
After substituting this equation into \rf{23} we obtain
%%%%%
\ba{30} &{}&6\dot{\Psi}=-\frac{1}{\dot{F}}\Phi\left[6H\dot{F}-FR+12FH^2\right]\nn\\
&-&\frac{1}{\dot{F}}\Psi\left[FR-12H^2F\right]
\nn\\
&+&\frac{2}{\dot{F}}\cfrac{F}{a^2}\left(\triangle\Phi+\triangle\Psi\right)
-\frac{2}{\dot{F}}\kappa^2\delta\varepsilon+\frac{6}{\dot{F}}H\frac{\kappa^2}{a^2}\Xi\,
. \ea
%%%%%

In addition, we substitute \rf{30} into \rf{23} and after some algebra obtain
%%%%%
\ba{34} &{}& 6\dot{\Phi}=-\Phi\left[12\cfrac{\dot{F}}{F}+\frac{FR}{\dot{F}}-12\frac{H^2F}{\dot{F}}\right]\nn \\
&+&\Psi\left[6\cfrac{\dot{F}}{F}
-6H+\frac{FR}{\dot{F}}-12\frac{H^2F}{\dot{F}}\right]\\
&-&\frac{2}{\dot{F}}\cfrac{F}{a^2}\left(\triangle\Phi+\triangle\Psi\right)-\left[\frac{6}{F}+\frac{6H}{\dot{F}}\right]\frac{\kappa^2}{a^2}\Xi
+\frac{2}{\dot{F}}\kappa^2\delta\varepsilon\, .\nn \ea
%%%%%

\vspace{1mm}

We have Eq.~\rf{7} which we have not used yet. Substituting Eqs. \rf{30} and \rf{34} into this equation, we arrive, after rather tedious calculations, at a short
and simple equation:
%%%%%%
\ba{40} &&\left[6FH^2+3\dot{H}F-\frac{1}{2}FR\right]\left(\Phi-\Psi\right)=0\, , \ea
%%%%%%
and this is an identity due to Eq.~\rf{2.7}. This confirms the consistency of our equations \rf{30} and \rf{34}. Additionally, Eqs.~\rf{15} and \rf{18} are
satisfied for $\dot\Psi$ and $\dot\Phi$ given by \rf{30} and \rf{34}. This can be proved after rather long algebra with the help of Eq.~\rf{2.17} for
$\delta\varepsilon$, the continuity equations $\dot\rho =-\nabla \left(\sum_p \rho_p {\mathbf{v}}_p\right)$ and $\dot\rho=-\nabla \left( \rho \mathbf{v}\right)$
(for the discrete and continuous inhomogeneities, respectively), the equation of motion $\dot\Xi=-H\Xi-{\overline\rho}\Phi/a$ (for both types of inhomogeneities),
Eqs. \rf{2.22} or \rf{2.24} and the background equations \rf{2.6} (where $\bar P=0$) and \rf{2.7}. Therefore, Eqs.~\rf{30} and \rf{34} are the only independent
differential equations for the potentials $\Psi$ and $\Phi$. These master equations can be rewritten in the final form:
%%%%%
%%%%%
\ba{107} &{}& \dot{\Psi} -\frac{1}{3}\cfrac{F}{\dot{F}a^2}\left(\triangle\Phi+\triangle\Psi\right)
+\Phi\left[H-\frac{F\dot{H}}{\dot{F}}\right] \nn\\
&+&\Psi\left[\frac{F\dot{H}}{\dot{F}}+\frac{\kappa^2\bar\rho}{\dot{F}a^3}\right]=-\frac{1}{3}\frac{\kappa^2\delta\rho}{\dot{F}a^3}
+\frac{H}{\dot{F}}\frac{\kappa^2}{a^2}\Xi\,  \ea
%%%%%
and
%%%%%
\ba{106} &{}& \dot{\Phi}+\frac{1}{3}\cfrac{F}{\dot{F}a^2}\left(\triangle\Phi+\triangle\Psi\right)-
\Psi\left[\cfrac{\dot{F}}{F}-H+\frac{F\dot{H}}{\dot{F}}+\frac{\kappa^2\bar\rho}{\dot{F}a^3}\right] \nn\\
&+&\Phi\left[2\cfrac{\dot{F}}{F}+\frac{F\dot{H}}{\dot{F}}\right]= \frac{1}{3}\frac{\kappa^2\delta\rho}{\dot{F}a^3}
-\left[\frac{1}{F}+\frac{H}{\dot{F}}\right]\frac{\kappa^2}{a^2}\Xi \, . \ea
%%%%%
%%%%%

\section{Conclusion}

\setcounter{equation}{0}

In this paper, we have considered the theory of scalar perturbations for nonlinear $f(R)$ models. The initial rather complicated system of six equations
\rf{1}--\rf{6} obtained previously in a number of papers (see, e.g., \cite{DeFelice:2010aj}) has been reduced to two independent differential master equations
\rf{107} and \rf{106} for two scalar potentials $\Phi$ and $\Psi$. It can be easily verified that in the linear model case $f(R)=R$, where $\Phi=\Psi$, both of
these equations are reduced to the following one:
%%%%%%%%
\be{41}
\triangle\Phi - \frac{3\kappa^2\bar\rho}{2a}\Phi
= \frac{\kappa^2}{2a}\delta\rho -\frac{3\kappa^2}{2}H\Xi\, ,
\ee
%%%%%%%%
where the time derivative disappears.
%and, second, the energy density fluctuation $\delta\varepsilon$ is given by Eq.~\rf{2.17} with an explicit dependence
%on the potential $\Phi$.
This equation exactly coincides with the one in papers \cite{Eingorn:2015hza,Eingorn:2016kdt,Eingorn:2015yra}. It is the Helmholtz equation (not the Poisson
equation), resulting in the Yukawa-type screening of the gravitational potential produced by matter inhomogeneities
\cite{Eingorn:2015hza,Eingorn:2016kdt,Eingorn:2015yra,Eingorn:2017}. This effect is called cosmic screening. The nonzero cosmological background (i.e., $\bar\rho
\neq 0$) is responsible for this effect. In the present paper we have applied the similar approach. Unfortunately, in the case of nonlinear $f(R)$ models we
cannot exclude the time derivatives from Eqs. \rf{107} and \rf{106} to get the pure Helmholtz-type equations. However, we have performed the full relativistic
analysis of the first-order theory of scalar perturbations for arbitrary nonlinear $f(R)$ models. In the literature, among such models, special attention is drawn
to the models that pass to the $\Lambda$CDM model. This can occur both at the early and late stages of the Universe evolution. In the early Universe such
transition is necessary to reproduce the conventional matter era \cite{Peel:2018aly,Nunes:2016drj}. On the other hand, in the late Universe the transition to the
de Sitter stage takes place in a number of viable models which possess the stable de Sitter points
\cite{DeFelice:2010aj,Capozziello:2018oiw,Eingorn:2014laa,Cognola:2005de,Cognola:2008zp,Bamba:2010iy,Motohashi:2011wy,Jaime:2013zwa}. Recently, the stable and unstable de Sitter stages for general $f(R)$ models were derived in \cite{Odintsov:2017tbc}. It is worth noting that nonlinear gravitational models which can unify the early and late stages of acceleration through the stage of dominance of matter were considered in \cite{Nojiri:2003ft,Nojiri:2006gh}.

As we have already mentioned above (see the text after Eq.~\rf{6}), our equations are applicable for $f(R)$ models at all spatial scales as long as the
approximation $\delta R/\bar R \ll 1$ works{\footnote{Our investigation was based on Eqs. \rf{1}--\rf{6} taken from the review \cite{DeFelice:2010aj}. The form of
these equations (see, in particular, \rf{6}) clearly demonstrates that they are valid under this approximation. Most of the research in the literature devoted to
the scalar perturbations in $f(R)$ models is based on this system of equations.}}. From this point of view, a particular interest is taken in a subclass of $f(R)$
models in which this approximation still works at the stage of transition from the $f(R)$  model to the $\Lambda$CDM model. The theory of scalar perturbations for
the $\Lambda$CDM model within the cosmic screening approach was investigated in detail in our papers \cite{Eingorn:2015hza,Eingorn:2016kdt,Eingorn:2015yra}.
Therefore, for such a subclass of $f(R)$ models, solutions of the equations for the scalar perturbations obtained in the present paper for the $f(R)$ models and
in the papers \cite{Eingorn:2015hza,Eingorn:2016kdt,Eingorn:2015yra} for the $\Lambda$CDM model can be matched to each other in the stage of transition. Hence, we
can continuously describe the large scale structure formation both on $f(R)$ and  $\Lambda$CDM stages.
%The obtained master equations for the potentials $\Phi$ and $\Psi$ are useful to create initial conditions for the numerical
%simulation of %the large scale structure of the Universe in $f(R)$ models.

%%%%%%%%%%%%%%%%%%%%%%%%%%%%%%%%%%%%%%%%%%%%%%%%%%%%%%%%%%%%
%%%%%%%%%%%%%%%%%%%%%%%%%%%%%%%%%%%%%%%%%%%%%%%%%%%%%%%%%%%%%%%%

%%%%%%%%%%%%%%%%%%%%%%%%%%%%%%%%%%%%%%%%%%%%%%%%%%%%%%%%%%%%%%%%%%%%%%%%%%%%%%%%
%%%%%%%%%%%%%%%%%%%%%%%%%%%%%%%%%%%%%%%%%%%%%%%%%%%%%%%%%%%%%%%%%%%%%%%%%%%%%%%%

\section*{Acknowledgements}

\"{O}A  acknowledges support by the Distinguished Young Scientist Award BAGEP of the Science Academy. AZ  acknowledges financial support from The Scientific and
Technological Research Council of Turkey (TUBITAK) in the scheme of Fellowships for Visiting Scientists and Scientists on Sabbatical Leave (BIDEB 2221). AZ also
acknowledges the hospitality of \.{I}stanbul Technical University (ITU) where parts of this work were carried out. The work of RB was partially supported by the
EMJMD Student Scholarship from Erasmus\,+\,: Erasmus Mundus Joint Master Degree programme AstroMundus in Astrophysics.

%%%%%%%%%%%%%%%%%%%%%%%%%%%%%%%%%%%%%%%%%%%%%%%%%%%%%%%%%%%%%%%%%%%%%%%%%%%%%%%%%%%%%%%
%%%%%%%%%%%%%%%%%%%%%%%%%%%%%%%%%%%%%%%%%%%%%%%%%%%%%%%%%%%%%%%%%%%%%%%%%%%%%%%%%%%%%%%

\end{document}